\documentclass[onecolumn,pre,aps,amsmath,amssymb,showpacs,superscriptaddress]{revtex4-2}
\usepackage{graphicx} 
\usepackage{amsmath}
\usepackage{amssymb}
\usepackage{color}
\usepackage{bm}
\usepackage{blindtext}
\usepackage{enumerate}
\usepackage{enumitem}
\usepackage{psfrag}
\usepackage{scalerel}

\def\tF{{\scaleto{F}{3.5pt}}}
\def\x{{\bm x}}

\def\u{{\bm u}}

\def\W{{\bm W}}

\def\bxi{{\bm{\xi}}}
\def\bpsi{{\bm \psi}}

\def\calF{\bm{\mathcal{F}}}
\def\F{{{\bm F}}}
\def\n{{{\bm n}}}
\def\d{{{\mbox{d}}}}

\usepackage{color}
\usepackage{natbib}

\setlist[description]{font=\normalfont\space}

\begin{document}
\title{Kinetic representation of the unified gas-kinetic wave-particle method and beyond}
  \author{Zhaoli Guo}
\email[Email:]{zlguo@mail.hust.edu.cn}
 \affiliation{Institute of Multidisciplinary Research for Mathematics and Applied Science, Huazhong University of Science and Technology, Wuhan 430074, China}
\affiliation{State Key Laboratory of Coal Combustion, Huazhong University of Science and Technology, Wuhan 430074, China}
 
\author{Yajun Zhu}
\email[Email:]{mazhuyajun@ust.hk}
\affiliation{Department of Mathematics, The Hong Kong University of Science and Technology, Clear Water Bay, Hong Kong, China}

\author{Kun Xu}
\email[Email:]{makxu@ust.hk}
\affiliation{Department of Mathematics, The Hong Kong University of Science and Technology, Clear Water Bay, Hong Kong, China}

\begin{abstract}
The unified gas-kinetic wave-particle (UGKWP) method is a hybrid method for multiscale flow simulations, in which the contributions to the whole gas evolution from deterministic hydrodynamic wave and stochastic particle transport are combined simultaneously. Originally, the UGKWP method was developed as a direct modeling approach at discrete level. In this work, we revisit the time evolution of each part of the involved simulation particles and wave molecules in UGKWP, and present the corresponding kinetic equations. The resultant kinetic system can be viewed as a collision decomposition of the original kinetic equation, which can serve as a basis for developing other kinetic methods for flows in all flow regimes.
\end{abstract}
\maketitle

\section{Introduction}
It is a challenging task to simulate gas flows involving different flow regimes in which the Knudsen number $\epsilon$ can vary in a wide range. Classical computational fluid dynamics (CFD) techniques based on Euler or Navier--Stokes equations are limited to continuum flows, while stochastic particle methods, such as the Direct Simulation Monte Carlo (DSMC) method, are mainly suitable for high-speed rarefied flows but usually encounter difficulties for continuum flows.

In recent years, a variety of deterministic discrete-velocity methods (DVMs) based on gas kinetic theory have been developed for simulating multiscale flows \cite{ref:Review14,ref:Luc2014Rev}. Particularly, kinetic schemes with asymptotic preserving (AP) properties have received much attention due to their capability in capturing the hydrodynamic behaviors in the limit of $\epsilon \to 0$ without resolving kinetic scales \cite{ref:AP_Rev_2012,ref:AP_Rev_2017}. A unified preserving (UP) concept was further proposed recently \cite{ref:UP2023}, which can be used to assess the order of asymptotics of a kinetic scheme for small but finite $\epsilon$ (i.e., $0<\epsilon\ll 1$). The unified gas-kinetic scheme (UGKS) \cite{ref:UGKS} and discrete unified gas-kinetic scheme (DUGKS) \cite{ref:DUGKS13} are two typical kinetic methods with UP properties, which have found wide applications in multiscale flow simulations \cite{ref:XuBook,ref:DUGKS_Rev}.

Although the deterministic DVMs with AP or UP properties can avoid limitations on mesh size and time step by kinetic scales, a large number of discrete velocities are required for flows involving multiple flow regimes ( particularly for high Mach number flows), leading to rather expensive memory and computational costs. Recently, a hybrid approach, named as unified gas-kinetic wave-particle (UGKWP) method, was developed to combine the advantages of stochastic particle method and the UGKS \cite{ref:WP_LiuJCP2020}. In UGKWP, the gas system is represented as a two-phase system, i.e., continuum hydrodynamic wave phase composed of gas molecules and discrete particle phase composed of simulation particles. The hydrodynamic wave reflects the collective dynamics of molecules undergoing intensive collisions, and describes the evolution of the near-equilibrium state; on the other hand, the discrete particle phase is composed of a number of simulation particles representing a cluster of gas molecules, which models the evolution of non-equilibrium states. The interaction between the two phases are realized by absorbing collided particles into wave and generating free transport particles from wave.

In the UGKWP method, particle transport is tracked directly such that no discrete particle velocity space is required, which significantly reduces the memory and computational costs. On the other hand, the hydrodynamic wave is captured from the equilibrium state deterministically. The contributions to the gas evolution from the wave and particle phases depend on the local cell Knudsen number, such that the transport of discrete particles dominates the solution in rarefied regime, while the hydrodynamic wave dominates the solution in continuum regime. Additionally, the number of particles changes adaptively with the local cell Knudsen number. This suggests that the particle phase and wave phase become negligible in the continuum and free molecular limits, respectively. Consequently, the UGKWP can serve as an efficient method for multiscale flow simulations.

The evolution of the entire gas system in UGKWP is described by the dynamics of simulation particles and hydrodynamic wave molecules, during which the two phases interact with each other. It should be noted, however, that UGKWP is designed at discrete level as a direct modeling method \cite{ref:XuBook}. In this work, we aim to present the corresponding kinetic equations for different types of particles/molecules based on their underlying dynamics. The resultant system can provide a theoretical basis for better understanding the physics of UGKWP and developing new numerical methods. 

The rest of this article is organized as follows. In Sec. II, we outline the kinetic equation based on which the UGKWP is developed, and discuss the solution structure as a superposition of uncollided and collided molecules. In Sec. III, we provide a brief review of the UGKWP method. The corresponding kinetic representation of the UGKWP method is presented in Sec. IV, together with an asymptotic analysis of the kinetic system and a revisit of UGKWP. In Sec. V, two other kinetic schemes are reinterpreted to demonstrate their connections with the kinetic system, and a brief summary is provided in the last section.

\section{BGK model and the solution structure}
\label{sec:BGK}
The UGKWP is constructed based on certain relaxation models of the Boltzmann equation. Without loss of generality, we consider the Bhatnagar-Gross-Krook (BGK) model for monatomic gases \cite{ref:BGK1954},
\begin{equation}
\label{eq:BGK}
\partial_t f +\bxi\cdot \nabla f = Q(f) \equiv -\dfrac{1}{\tau}\left(f-g\right),
\end{equation}
where $f(\x,\bxi,t)$ is the distribution function for gas molecules moving with velocity $\bxi$ at position $\x$ and time $t$, $Q$ is the collision operatior, $\tau$ is the relaxation time, and $g$ is the equilibrium distribution function defined by
\begin{equation}
    g(\x,\bxi,t)=\mathcal{E}(\W(\x,t),\bxi)\equiv \dfrac{\rho}{(2\pi RT)^{D/2}}\exp\left(-\dfrac{|\bxi-\u|^2}{2RT}\right).
\end{equation}
Here $D$ is the spatial dimension, $\W=\rho(1,\u,E)$ is the conservative hydrodynamic variables, with density $\rho$, velocity $\u$, and total energy $E=(|\u|^2+DRT)/2$ (with $R$ as the specific gas constant and $T$ as the temperature). $\W$ is determined by the conservative moments of the distribution function, i.e.,  $\W=\int{\bpsi(\bxi) f \d\bxi}$ with $\bpsi=(1,\bxi,\xi^2/2)$. Taking the conservative moments of Eq. \eqref{eq:BGK}, we can derive the macroscopic conservative equations
\begin{equation}
\label{eq:Wt}
    \partial_t \W + \nabla\cdot \calF =0,
\end{equation}
where $\calF=\int \bxi\bpsi f \d \bxi$ is the macroscopic hydrodynamic flux.

The BGK equation \eqref{eq:BGK} admits a formal analytical solution,
\begin{equation}
	\label{eq:BGK-solution}
	f(\x,\bxi,t)=\underbrace{e^{-\nu(\x,t)} f(\x-\bxi (t-t_0),\bxi,t_0)}_{f_u(\x,\bxi,t)} + \underbrace{ \int_{t_0}^t {\left[\dfrac{1}{\tau(\x',t')} e^{-(\nu(\x,t)-\nu(\x,t'))} g(\x',\bxi, t') \right]\d t'}}_{f_c(\x,\bxi,t)},
\end{equation}
where $t_0$ is the initial time, $\x'=\x-\bxi(t-t')$, and $ \nu(\x,t)=\int_{t_0}^t{[\tau\left(\x-\xi(t-s),s\right)]^{-1}\d s}$. Equation \eqref{eq:BGK-solution}  demonstrates the structure of the solution of the BGK equation: $f_u(\x,\bxi,t)$ represents the remaining uncollided molecules transported from $t_0$ to $t$; while $f_c(\x,\bxi,t)$ represents collided molecules undergoing intensive collisions. It is noted that if $\tau$ is a (local) constant, then $\nu(\x,t)=(t-t_0)/\tau$, and the formal solution reduces to 
\begin{equation}
	\label{eq:BGK-solution-Const}
	f(\x,\bxi,t)=e^{-(t-t_0)/\tau} f(\x-\bxi(t-t_0),\bxi, t_0) + \dfrac{1}{\tau} \int_{t_0}^t {e^{-(t-t')/\tau} g(\x',\bxi, t') \d t'}.
\end{equation}
We emphasize here that $f(\x-\bxi (t-t_0),\bxi,t_0)$ represents all molecules at the initial time $t_0$, which can either transport freely (collisionless molecules) or undergo collisions (collisional molecules) at later times.

\begin{figure}
	\centering
	\includegraphics[width=0.6\textwidth]{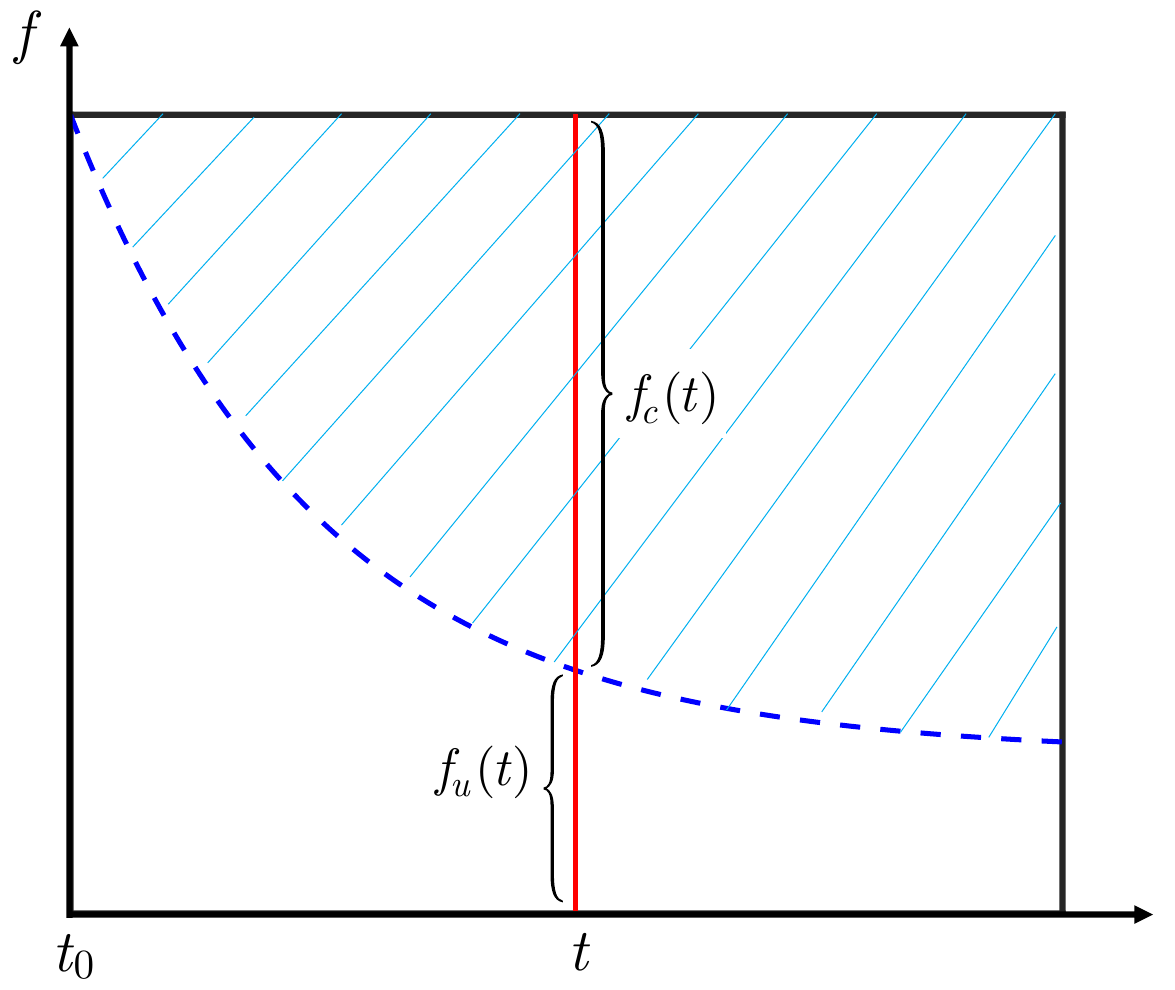}    
	\caption{\label{fig:timeFUC} Time evolutions of the distribution functions of collided and uncollided molecules. Dashed line: The first-collision time of uncollided molecules; Shading region: collided molecules in continuum wave phase; Blank region: Uncollided molecules in discrete phase.}
\end{figure}

The analytical solution \eqref{eq:BGK-solution} suggests that the distribution function can be decomposed into two parts, i.e.,  $f=f_u+f_c$. This decomposition indicates that the gas system can be considered as a ``two-phase" system: the uncollided molecules forms a {\em discrete  phase} since they are independent with each other, while the collided molecules constitute a {\em continuum wave phase} since they undergo intensive collisions, which is also evident that $f_c$ depends only on the equilibrium distribution function $g$. Furthermore, the kinetic equations and corresponding initial conditions for both phases can be derived from the definitions of $f_u$ and $f_c$, 
\begin{subequations}
\label{eq:BGK_2Popolation}
\begin{equation}
	\label{eq:BGK_fu}
	\left\{
	\begin{aligned}
		&\partial_t f_u +\bxi\cdot \nabla f_u = -\dfrac{1}{\tau} f_u, \\
		& f_u(\x,\bxi,t_0) = f(\x,\bxi, t_0),
	\end{aligned}
	\right.
\end{equation}
\begin{equation}
	\label{eq:BGK_fc}
	\left\{
	\begin{aligned}
		&\partial_t f_c +\bxi\cdot \nabla f_c = -\dfrac{1}{\tau} (f_c-g), \\
		& f_c(\x,\bxi,t_0) = 0.
	\end{aligned}
	\right.
\end{equation}
\end{subequations}
The initial conditions, $f_u(\x,\xi,t_0)=f(\x,\xi,t_0)$ and $f_c(\x,\xi,t_0)=0$, indicate that all molecules are initially uncollided. Furthermore, the relaxation terms in Eqs. \eqref{eq:BGK_fu} and \eqref{eq:BGK_fc} suggest that the number of uncollided molecules decays exponentially with time, and collided molecules are generated due to collisions among all molecules. From a particle perspective, the probability of a molecule experiencing no collisions in time period $[t_0,t]$ is $p=e^{-\nu(t)}$. The time evolutions of $f_u$ and $f_c$ are sketched in Fig. \ref{fig:timeFUC}, where the dashed line represents the first collision time (FCT) of molecules. A molecule in discrete phase transports freely until it encounters its first collision with other molecules at FCT, at which point it transits to the collided state in the wave phase.

With the decomposition $f=f_u+f_c$, the macroscopic variables $\W$ and the flux $\calF$ can also be decomposed into two parts accordingly: $\W=\W_u+\W_c$ and $\calF=\calF_u+\calF_c$, where
\begin{equation} 
	\W_\alpha(\x,t)=\int{\left[\bpsi(\bxi) f_\alpha(\x,\bxi,t)\right] \d \bxi}, \quad \calF_\alpha(\x,t)=\int\left[\bxi\bpsi(\bxi) f_\alpha (\x,\bxi,t)\right]  \d \bxi,\quad \alpha =u, c.
\end{equation}
The governing equations for $\W_u$ and $\W_c$ can be obtained by taking the moments of Eqs. \eqref{eq:BGK_fu} and Eqs. \eqref{eq:BGK_fc}, respectively, namely, 
\begin{subequations}
	\begin{equation}
		\label{eq:BGK_Wu}
		\left\{
		\begin{aligned}
			&\partial_t \W_u + \nabla\cdot \calF_u =-\dfrac{1}{\tau}\W_u, \\
			& \W_u(\x,t_0)=\W(\x,t_0),
		\end{aligned}
		\right.
	\end{equation}
	\begin{equation}
		\label{eq:BGK_Wc}
		\left\{
		\begin{aligned}
			&\partial_t \W_c + \nabla\cdot \calF_c  = \dfrac{1}{\tau}\W_u,\\
			& \W_c(\x,t_0)=0.
		\end{aligned}
		\right.
	\end{equation}
\end{subequations}
The source terms on the right-hand side of the above two equations reflect the interaction between the discrete and continuum phases.

Equations \eqref{eq:BGK_fu} and \eqref{eq:BGK_fc} can be viewed as a decomposition of the original BGK equation \eqref{eq:BGK} based on molecular collision behaviors. Actually, if $f_u$ and $f_c$ are solutions of Eqs. \eqref{eq:BGK_fu} and \eqref{eq:BGK_fc}, respectively, then $f=f_u+f_c$ is the solution of the BGK equation \eqref{eq:BGK}. From this viewpoint, this decomposed two-population system is equivalent to the BGK equation \eqref{eq:BGK}, meaning that no splitting errors exist in the decomposition system. This contrasts with the classical time-splitting method, which separates the kinetic equation into a pure convection equation and a pure collision equation such that time-error is introduced. 

It should be pointed out that the idea of uncollided-collided decomposition was first proposed in the ``first collision source" method for solving neutron transport equation with a source term \cite{ref:Coll-Hybrid1971}. By solving the uncollided particle equation with deterministic or stochastic methods, ray effects can be effectively reduced \cite{ref:Coll-Hybrid1971,ref:Coll-Hybrid1985,ref:Coll-Hybrid2013}. Recently, this method was extended to solve the Boltzmann-BGK equation with the asymptotic Euler limit \cite{ref:Coll-Hybrid-BGK2024}. However, in previous studies, the kinetic equations for $f_u$ and $f_c$, as well as the initial conditions, are proposed intuitively. In contrast, in the present work we derived the kinetic equations and initial conditions rigorously from the analytical solution. An additional advantage of the present approach is that it clearly identifies the structure of solution and reveals the time evolution of each part, as demonstrated in Fig. \ref{fig:timeFUC}.

The two-population system given by Eqs. \eqref{eq:BGK_fu} and \eqref{eq:BGK_fc} also offers some advantages for designing numerical methods. Particularly, Eq. \eqref{eq:BGK_fu} is linear, and its solution can be obtained analytically as defined in \eqref{eq:BGK-solution}. On the other hand, Eq. \eqref{eq:BGK_fc} is non-linear, but $f_c(\x,\bxi, t)$ is fully determined by the equilibrium distribution $g$ along the characteristic line, which means it can be calculated directly from the macroscopic variables $\W$. Consequently, the two sub-kinetic equations can be solved using different numerical methods, leading to efficient hybrid methods with desired properties. 

\section{UGKWP method}
The UGKWP method was developed based on the unified gas kinetic particle (UGKP) method, which is a hybrid particle-continuum method \cite{ref:KP2018,ref:WP_LiuJCP2020}. In this section, we will first provide a brief introduction to the UGKP method and then present the main details of the UGKWP method. 

\subsection{UGKP method}
The UGKP method is a finite-volume discretizations of Eq. \eqref{eq:Wt}, where the numerical flux across each cell interface is reconstructed based on the analytical solution \eqref{eq:BGK-solution}. The physical space is discretized into a number of control volumes (cells), and the update of cell averaged macroscopic variables $\W$ in a cell $V_i$ from time $t_n$ to $t_{n+1}=t_n+\Delta t$ is given by 
\begin{equation}
\label{eq:discrete_W}
    \W^{n+1}_i-\W^n_i + \dfrac{1}{|V_i|}\sum_{j\in N(i)} \bm{F}_{ij} |S_{ij}| = 0,
\end{equation}
where $|V_i|$ is the volume of cell $V_i$, $N(i)$ is the set of interface adjacent neighboring cells of $V_i$, $S_{ij}$ is the interface between cells $V_i$ and $V_j$, and $|S_{ij}|$ is the surface area. Additionally, $\W_i$ is the volume-averaged macroscopic variables, 
\begin{equation}
    \W_i^n=\dfrac{1}{|V_i|}\int_{V_i}{\W(\x,t_n) \d\x},
\end{equation}
and $\bm{F}_{ij}$ is the flux across the interface $S_{ij}$,
\begin{equation}
    \bm{F}_{ij}=\int_{t_n}^{t_{n+1}}{\int (\bxi\cdot\n_{ij})\bpsi(\bxi) f(\x_{ij},\bxi,t) \d \bxi\d t},
\end{equation}
where $\n_{ij}$ the unit normal vector of interface $S_{ij}$ pointing from cell $V_i$ to cell $V_j$, and $\x_{ij}$ is the center of $S_{ij}$.

The flux $\bm{F}_{ij}$ depends on the instantaneous distribution function $f(\x_{ij},\bxi,t)$, which can be obtained from the formal analytical solution of the BGK equation. Here we assume that the relaxation time $\tau$ is a local constant during $t_n\le t < t_{n+1}$ around $\x_{ij}$, then we can adopt the formulation in Eq. \eqref{eq:BGK-solution-Const} by setting $t_0\to t_n$,
\begin{equation}
	\label{eq:IntSolution}
	f(\x_{ij},\bxi,t)=\underbrace{e^{-(t-t_n)/\tau} f(\x-\bxi(t-t_n),\bxi, t_0)}_{f_u(\x_{ij},\bxi,t)} + \underbrace{\dfrac{1}{\tau} \int_{t_n}^t {e^{-(t-t')/\tau} g(\x',\bxi, t') \d t'}}_{f_c(\x_{ij},\bxi, t)}, \quad t_n\le t <t_{n+1}.
\end{equation}
Accordingly, the interface flux can be decomposed into two parts, $\bm{F}_{ij}=\bm{F}_{ij}^u+\bm{F}_{ij}^c$, where
\begin{equation} 
 \bm{F}_{ij}^\alpha=\int_{t_n}^{t_{n+1}}{\int (\bxi\cdot\n_{ij})\bpsi(\bxi) f_\alpha (\x_{ij},\bxi,t) \d \bxi\d t},\quad \alpha =u, c.
\end{equation}

It is noted that $f_c$ depends only on the hydrodynamic variables $\W$; therefore, $\bm{F}_{ij}^c$ can be evaluated explicitly from $\W$. In the UGKP method, a first-order Taylor expansion of $g$ is adopted in the time integral in Eq. \eqref{eq:BGK-solution-Const}, 
\begin{equation}
g(\x'_{ij},\bxi,t')=g_{ij}^n-(t-t')\bxi\cdot\nabla g_{ij}^n+(t'-t_n)\partial_t g_{ij}^n,    
\end{equation}
where $g_{ij}^n=g(\x_{ij}, \bxi, t_n)$. Then the collided distribution function $f_c$ can be approximated as
\begin{equation}
f_c(\x_{ij},\bxi,t)= \dfrac{1}{\tau} \int_{t_n}^t {e^{-(t-t')/\tau} g(\x'_{ij},\bxi, t') \d t'}=c_1 g_{ij}^n +c_2 \bxi\cdot\nabla g_{ij}^n + c_3 \partial_t g_{ij}^n,
\end{equation}
where $\x'_{ij}=\x_{ij}-\bxi(t-t')$, and 
\begin{equation}
\label{eq:c1-3}
c_1=1-e^{-s/\tau}, \quad c_2=s-(s+\tau) c_1, \quad c_3=s-\tau c_1, \quad s=t-t_n.
\end{equation}
Then the contribution from $f_c$ over the whole time step can be obtained as
\begin{equation}
	\label{eq:FijC}
    \bm{F}_{ij}^c=\int (\bxi\cdot\n_{ij})\bpsi(\bxi) \left[C_1 g_{ij}^n +C_2 \bxi\cdot\nabla g_{ij}^n + C_3 \partial_t g_{ij}^n\right] \d \bxi,
\end{equation}
where
\begin{equation}
C_1=\Delta t-\tau \left(1-e^{-\Delta t/\tau} \right), \quad C_2=\Delta t^2 -(2\tau+\Delta t) C_1, \quad C_3=\dfrac{\Delta t^2}{2}-\tau C_1.
\end{equation}

The flux from the uncollided molecules, $\bm{F}_{ij}^u$, is evaluated using a particle method in UGKP; in other words, the discrete phase is represented by simulation particles. The flow field at time $t_0$ is initialized as follows: from the given initial macroscopic variables $\W_i^0$ in cell $V_i$, a set of simulation particles $\{P_{ik}^0(m_k,\x_k,\bxi_k): k=1,\cdots, N_i^0 \}$ is sampled, and the particle $P_k$ is characterized by its mass $m_k$, position $\x_k$, and velocity $\bxi_k$. It is noted that each simulation particle represents a cluster of gas molecules. At the beginning of each time step, all particles are treated as uncollided ones, so $\W_{u,i}^0=\W_i^0$ and $\W_{c,i}^0=0$. Then evolution procedure from $t_n$ to $t_{n+1}$ can then be described as follows.
\begin{enumerate}[label={\arabic*.}]
	    \item Resampling: \\
	    As $n>0$, keep the remaining uncollided particles at $t_n^-$, and resample particles from the wave phase following the Maxwellian distribution $\mathcal{E}(\W_{c,i}(t_n^-),\bxi)$. This means that the whole flow field in $V_i$ is represented by $N_i^n$ simulation particles at the moment of $t_n^+$.
	    \item Classification:\\
          Classify the particles in $V_i$ into $N_{i,f}^n$ collisionless particles with probability $\beta=e^{-\Delta t/\tau}$ and $N_{i,c}^n$ collisional ones: Each particle ($P_k$) is assigned a first collision time $t_{k,c}=\min(-\tau\ln r, \Delta t)$, where $r$ is a uniform random number in $[0,1]$; then the particle is called as a ``collisionless particle" if $t_{kc}=\Delta t$, and as a``collisional particle" otherwise. Approximately, $N_{i,f}^n\approx \beta N_i^n$ and $N_{i,c}^n\approx (1-\beta) N_i^n$.
          
	    \item Transport:\\
	    Move all particles to new positions according to their velocities and first collision times, i.e.,  $\x_k^{n+1}=\x_k^n+\bxi_k t_{kc}$ for particle $P_k$. Then keep collisionless particles (those with $t_{kc}=\Delta t$) and delete collisional particles. The flux across interface $S_{ij}$ is measured by recording the particles passing through the interface,
    	\begin{equation}
    		\label{eq:FijU}
   	        \F_{ij}^u=\sum_{k=1}^{N_{ij}} {(\bxi_k\cdot\n_{ij})\bpsi_k m_k},
        \end{equation}
        where $\bpsi_k=(1,\bxi_k,\xi_k^2/2)$, and $N_{ij}$ is the number of particles passing through $S_{ij}$. 
        
	    \item Updating macroscopic variables: 
	    \begin{enumerate}[label={(\alph*)}]
	    		\item Calculate macroscopic variables of the particle phase, 
	    \begin{equation}
	    	\label{eq:Wui}
		    \W_{u,i}^{n+1}=\dfrac{1}{|V_i|}\sum_{k=1}^{N_i^{n+1}} {\bpsi_k m_k}
	    \end{equation}
	     where $N_i^{n+1}$ is the updated number of particles in the cell. 
   		\item Compute $\W_i^{n+1}$ according to Eq. \eqref{eq:discrete_W}, where the flux $\F_{ij}=\F_{ij}^u+\F_{ij}^c$ is given by Eqs. \eqref{eq:FijC} and \eqref{eq:FijU}.
	     \item Calculate macroscopic variables for the wave phase composed of collided particles, $\W_{c,i}^{n+1}=\W_i^{n+1}-\W_{u,i}^{n+1}$.
    \end{enumerate}
\end{enumerate}
\begin{figure}
	\centering
	\includegraphics[width=0.6\textwidth]{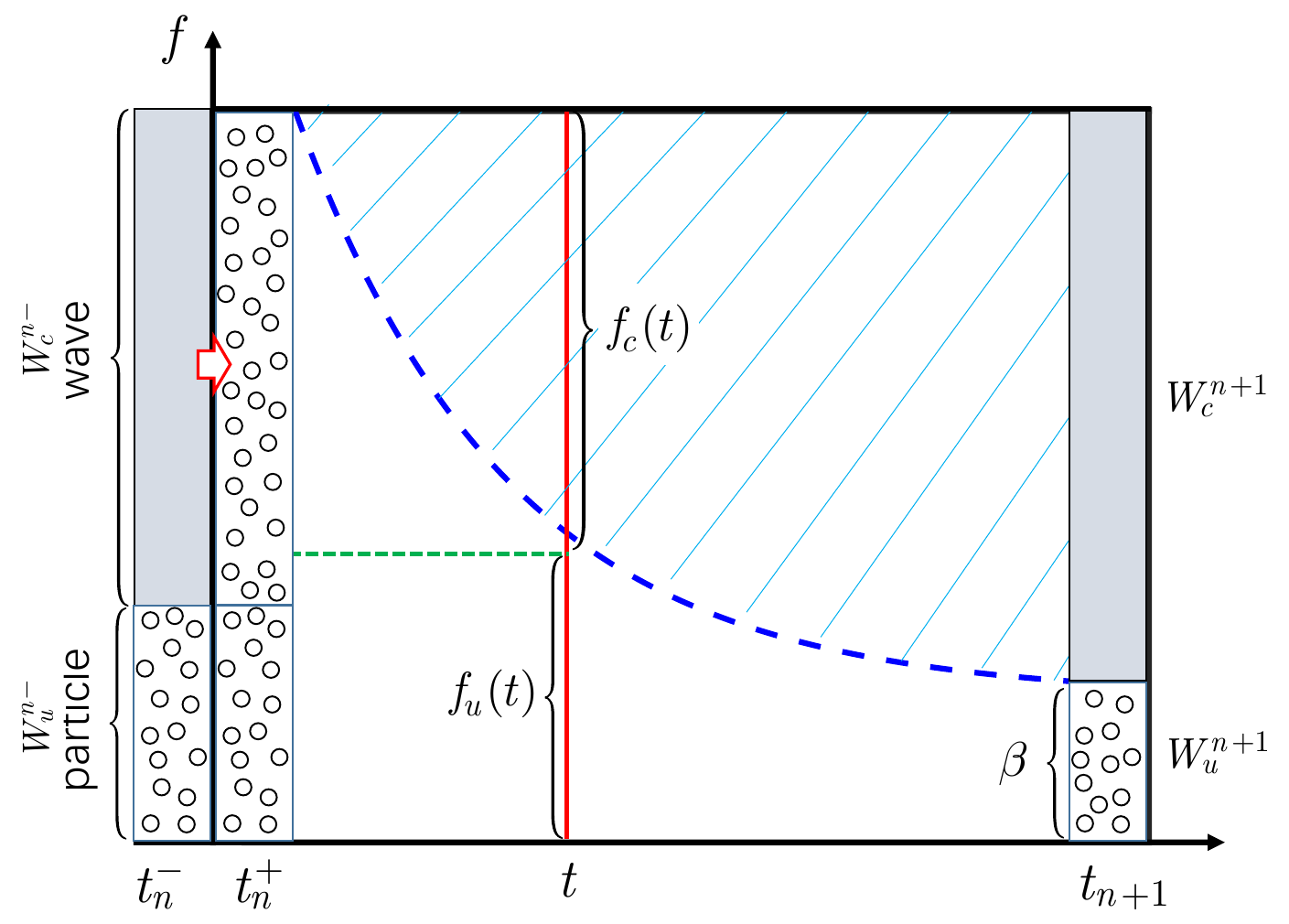}    
	\caption{\label{fig:UGKP} Particle evolution in UGKP. Gray: wave phase; Circle: particle phase; Blue dashed line: first collision time; Green dashed line: boundary for collided (above) and uncollided (below) particles between $t_n$ and $t$. $\beta=e^{-\Delta t/\tau}$ is the portion of all initial particles that survive at $t_{n+1}$.}
\end{figure}

The particle dynamics and changes of state in UGKP are illustrated in Fig. \ref{fig:UGKP}. The evolution procedureshows that two phases are involved in UGKP: the wave phase and the particle phase. The wave phase is composed of gas molecules and is continuum, while the particle phase consists of discrete simulation particles. Once an uncollided particle undergoes collisions with other particles, it will be deleted and absorbed into the wave phase. 

Furthermore, UGKP can also be reinterpreted as a hybrid method based on the decomposed two-population kinetic system described in Sec. II. Specifically, UGKP solves the following two-population kinetic system,
\begin{subequations}
\label{eq:KP_2Popolation}
\begin{equation}
	\label{eq:UGKP_fu}
	\left\{
	\begin{aligned}
		&\partial_t f_u +\bxi\cdot \nabla f_u = -\dfrac{1}{\tau} f_u, \quad t_n\le t<t_{n+1}\\
		& f_u(\x,\bxi,t_n^+) = f_u(\x,\bxi, t_n^-) + \mathcal{E}(\W_c(\x,t_n^-),\bxi),
	\end{aligned}
	\right.
\end{equation}
	\begin{equation}
	\label{eq:UGKP_fc}
	\left\{
	\begin{aligned}
		&\partial_t f_{c} +\bxi\cdot \nabla f_{c} = -\dfrac{1}{\tau}\left(f_{c}-g\right), \quad t_n < t \le t_{n+1}, \\
		& f_{c}(\x,\bxi,t_n^+) = 0,
	\end{aligned}
	\right.
\end{equation}
\end{subequations}
where $\mathcal{E}(\W_c(\x,t_n^-),\bxi)$ represents the resampled particles from the wave phase as illustrated in Fig. \ref{fig:UGKP}. Equation \eqref{eq:UGKP_fu} is solved using particle method at each time step with a re-initialized condition. On the other hand, the kinetic equation \eqref{eq:UGKP_fc} is solved deterministically. Particularly, since the collective dynamics of the collided particles behaves like hydrodynamic wave, it is not necessary to solve $f_c$ directly from the kinetic equation \eqref{eq:UGKP_fc}. Instead, UGKP solves effectively the following moment equation for collided particles within each time step, 
	\begin{equation}
	\label{eq:UGKP_Wc}
	\left\{
	\begin{aligned}
		&\partial_t \W_c + \nabla\cdot \calF_c  = \dfrac{1}{\tau}\W_u, \quad t_n\le t<t_{n+1}\\
		& \W_c(\x,t_n^+)=0,
	\end{aligned}
	\right.
\end{equation}
or equivalently the moment equation for the entire system,
	\begin{equation}
	\label{eq:UGKP_W}
	\left\{
	\begin{aligned}
		&\partial_t \W + \nabla\cdot \calF  = 0, \quad t_n\le t<t_{n+1}\\
		& \W(\x,t_n^+)=\W(\x,t_n^-).
	\end{aligned}
	\right.
\end{equation}

The reinterpretation of UGKP as a collision-based hybrid method provides a kinetic description of this method,  even though it was initially constructed following a direct modelling approach. Furthemore, UGKP gives an example how the two-population system described by Eq. \eqref{eq:BGK_2Popolation} can offer advantages over the original kinetic equation in designing numerical methods.

\subsection{UGKWP method}
The UGKWP method is an improvement of UGKP, which still represents the gas as a two-phase system composed of discrete particles and continuum waves. In the UGKP method, the wave phase $\W_c$ is fully resampled as discrete particles at the beginning of each time step ($t_n^+$). However, it is noted that only a portion of $\beta=e^{-\Delta t/\tau}$ of these particles survive to $t_{n+1}$. This implies that other resampled particles become collisional and get removed at $t_{n+1}$. Therefore, it is unnecessary to sample these collisional particles from the wave phase initially at the beginning of each time step. Based on this observation, the UGKP is improved to UGKWP by sampling only the collisionless particles at $t_n^+$. 

The first improvement lies in the initialization of the flow field at $t_0$. In UGKP, the gas system ($\W^0$) is fully represented as simulation particles. In contrast, UGKWP samples only a portion $\beta$ of $\W^0$ as particles, while the remaining $(1-\beta)\W^0$ remains at continuum wave state. Thus, $\W_u^0=\beta\W^0$ and $\W_c^0=(1-\beta)\W^0$, with the corresponding distribution functions $f_u(t_0)=\mathcal{E}(\W_u^0,\bxi)$ and $f_c(t_0)=\mathcal{E}(\W_c^0,\bxi)$. Consequently, the number of sampled particles in the initial step of UGKWP is less than that in UGKP, particularly for near continuum flows ($\beta\ll 1$).

The second improvement lies in the particle resampling from wave phase at later times. In UGKP, the entire wave phase $\W_c(t_n^-)$ is resampled as simulation particles that may either transport freely or encounter collisions. In contrast, UGKWP only samples collisionless particles that transport freely, which is proprotional to $\beta\W(t_n^-)$, in the whole time step are re-sampled. As a result, the gas system at $t_n^+$ contains both particle and wave phases, leading to a significant reduction in the number of simulation particles.

The time evolution of the wave phase also differs between UGKP and UGKWP. In UGKP, the wave phase is absent at $t_n^+$ and is generated from the removed collided particles at later time. On the other hand, in UGKWP both particle and wave phases exist at $t_n^+$, and additional wave is generated from the removed collided particles and molecules in both phases.

The detailed evolution procedure from $t_n$ to $t_{n+1}$ of UGKWP is listed as follows:
\begin{enumerate}[label={\arabic*.}]
\item Classification: \\
Classify the $N_i^{n}$ particles in cell $V_i$ into $N_{pf}$ free transport (collisionless) particles with probability $\beta=e^{-\Delta t/\tau}$ and $N_{c}=N_i^{n}-N_{pf}$ collisional ones. 
\item Resampling: \\
As $n>0$, resample $N_{wf}$ collisionless particles from the wave phase following the Maxwellian distribution $\mathcal{E}(\beta\W_{c,i}(t_n^-),\bxi)$. The total number of collisionless particles in $V_i$ is $N_f=N_{pf}+N_{wf}$ then.
\item
Particle transport:\\
Move all particles (including the resampled ones from wave phase) to new positions according to their velocities and flight time, then keep collisionless particles and delete collisional particles. The flux from particles, $\F_{ij}^p$, is measured by recording the particles passing through the interface, as given by Eq. \eqref{eq:FijU}. 
\item
\label{Flux}
Updating macroscopic variables:\\
   \begin{enumerate}[label={(\alph*)}]
      \item Update the macroscopic variables of the particle phase, $\W_{p,i}^{n+1}$, as given by Eq. \eqref{eq:Wui}.
\item Calculate the total macroscopic variables,
\begin{equation}
\label{eq:WP_tEvol}
\W_i^{n+1}=\W_i^n-\dfrac{1}{|V_i|}\sum_{j\in N(i)} \bm{F}^c_{ij} |S_{ij}|-\dfrac{1}{|V_i|}\sum_{j\in N(i)}\bm{F}_{ij}^{wr} |S_{ij}|-\dfrac{1}{|V_i|}\sum_{j\in N(i)}\bm{F}_{ij}^p|S_{ij}|,
\end{equation}
where $\F_{ij}^c$ is the flux contributed from the collided particles and collided wave molecules, $\F_{ij}^p$ is the flux from all simulation particles including the original ones and those resampled from wave phase, and $\F_{ij}^{wr}$ is the flux from the un-sampled collisional molecules of the wave phase in the process before their first collisions, which is given by
\begin{equation}
\label{eq:fwr}
    \F_{ij}^{wr} =\int_{t_n}^{t_{n+1}}{\int (\bxi\cdot\n_{ij})\bpsi(\bxi) \left[(e^{-(t-t_n)/\tau}-e^{-\Delta t/\tau})g_w(\x_{ij}-\bxi(t-t_n), \bxi, t_n^+)\right]\d \bxi\d t},
\end{equation}    
where $g_w(t_n^+)=\mathcal{E}(\W_c'(t_n^-))$ with $\W_c'=(\rho_c/\rho)\W$, namely, $g_w$ is the Maxwellian distribution with temperature and average velocity determined by the total macroscopic variables $\W$, but with the density replaced by $\rho_c$.
\item  
Calculate the macroscopic variables for the collided particles/molecules, $\W_{c,i}^{n+1}=\W_i^{n+1}-\W_{p,i}^{n+1}$.
  \end{enumerate}
\end{enumerate}

\begin{figure}
	\centering
	\includegraphics[width=0.6\textwidth]{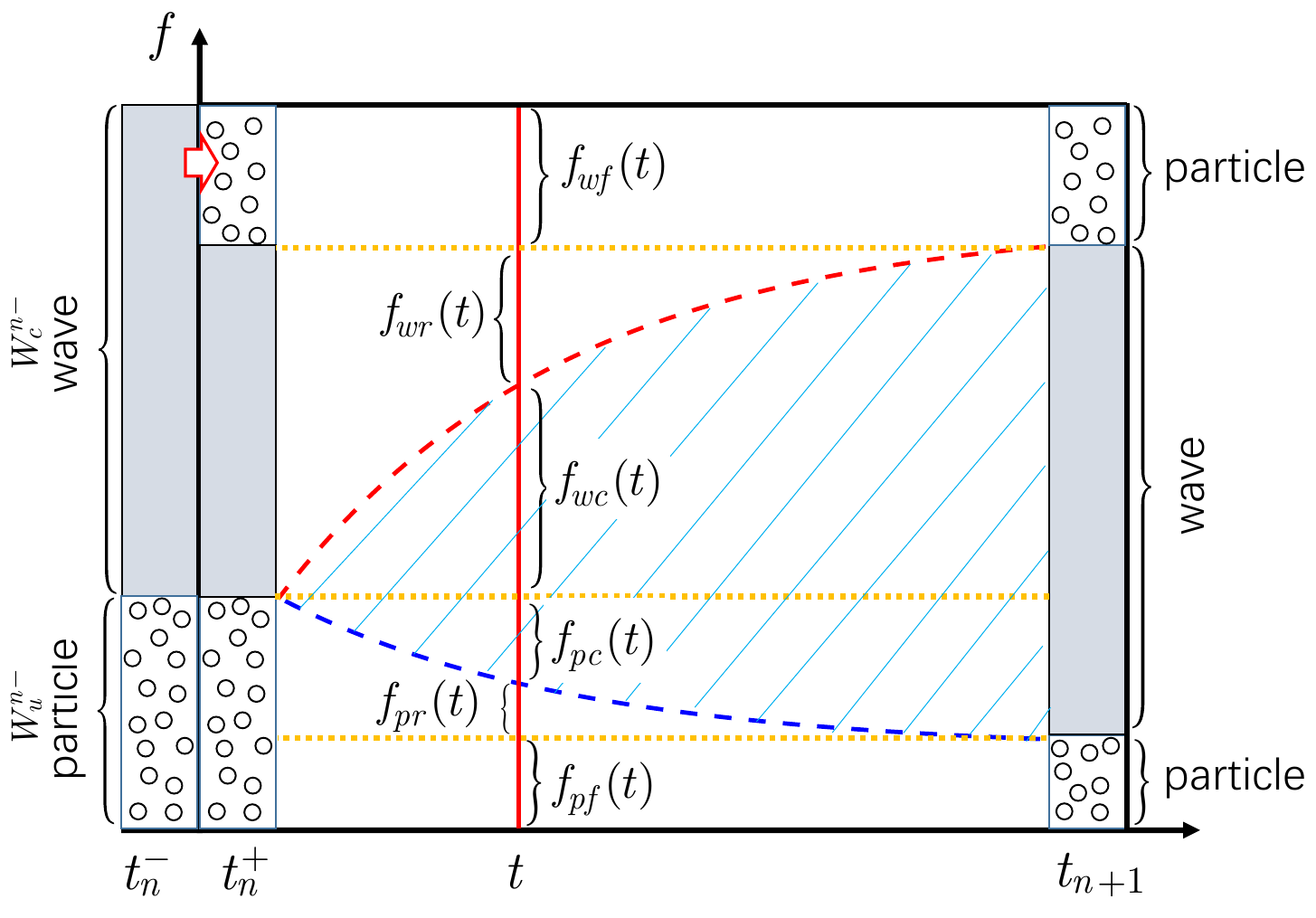}    
	\caption{\label{fig:UGKWP} Particle evolution in UGKWP. Gray: wave phase; Circle: particle phase; Blue dashed line: first collision time of particles; Red dashed line: first collision time of molecules in wave phase. Representations of the distribution functions: $f_{wf}$-- free transport wave molecules; $f_{wr}$-- collisional wave molecules that will experience collisions and get removed; $f_{wc}$-- collided wave molecules; $f_{pc}$-- collided particle; $f_{pr}$-- collisional particles that will experience collisions and get removed; $f_{pf}$-- free transport particles.}
\end{figure}

The dynamics of the particle and wave phases in UGKWP is illustrated in Fig. \ref{fig:UGKWP}. From the evolution procedure, we can see clearly the interaction between the wave phase and particle phase: simulation particles are resampled from wave phase at the beginning of each time step, and wave is generated from collisional particles. Furthermore, it can be found that the number of particles in UGKWP is proportional to the parameter $\beta=e^{-\Delta t/\tau}$. This suggests that there will be very few particles in cells where the flow is continuum ($\Delta t\gg\tau$), and the flow will be dominated by the hydrodynamic wave phase. Conversely, as $\Delta t\ll \tau$, the wave part becomes insignificant and the flow is dominated by the particle phase. Overall, the UGKWP method can serves as an efficient dynamical multiscale approach for flows ranging from continuum to free-molecular regimes.

\section{Kinetic representation of UGKWP}
\subsection{Kinetic equations for different types of simulation particles and wave molecules}
\label{sec:KineticSystem}
The construction of the interface flux in UGKWP is implemented at discrete level in a direct modelling manner \cite{ref:XuBook}. To provide a more clear understanding of the physical processes involved, we now present the kinetic equations for different types of particles and molecules in the discrete and continuum phases. 

As sketched in Fig. \ref{fig:UGKWP}, at time $t$ the entire system can be represented by various types of wave molecules and simulation particles. In other words, the distribution function $f$ can be decomposed into six parts, 
\begin{equation}
	f(t)=\underbrace{f_{pf}(t)+f_{pr}(t)+f_{pc}(t)}_{particle}+\underbrace{f_{wf}(t)+f_{wr}(t)+f_{wc}(t)}_{wave},
\end{equation}
where the sub-distribution functions for different particles/molecules are defined according to their phase states and collision behaviors, namely,
%
\begin{enumerate}[leftmargin=!,labelindent=2em,rightmargin=2em,label={}]
    \item[$f_{pf}$:]
       collisionless particles (free transport particles during the whole time step), $t_c\ge \Delta t$; 
    \item[$f_{pr}$:] 
       collisional particles (uncollided particles before the first collision, which will be removed due to collisions), $0<t_c< \Delta t$;
    \item[$f_{pc}$:] collided particles, $t_c=0$;
    \item[$f_{wf}$:] collisionless molecules (free transport wave molecules, resampled as simulation particles), $t_c\ge \Delta t$;
    \item[$f_{wr}$:] collisional molecules (uncollided wave molecules before the first collision, which will be removed due to collisions), $0<t_c< \Delta t$;
    \item[$f_{wc}$:] collided molecules in wave, $t_c=\Delta t$. 
\end{enumerate}

Here the subscripts $p$ and $w$ represent particle phase and wave phase, respectively, while $f$, $r$, and $c$ represent the different collision processes, respectively. It should be noted that $f_{pr}$ and $f_{wr}$ are distribution functions of the uncollided particles and wave molecules in the transport process before their first collisions, which will experience collisions and be removed after collisions. Actually, $f_{pr}$ and $f_{wr}$ are intermediate variables used to better demonstrate the decomposition of the entire distribution function and the transition process from discrete state to continuum state.
 
The kinetic equations for different parts can be obtained based on their time evolutions in UGKWP, as described below,
\begin{subequations}
	\label{eq:WP-6Parts}
\begin{equation}
	\label{eq:BGK_fpf}
	\left\{
	\begin{aligned}
		&\partial_t f_{pf} +\bxi\cdot \nabla f_{pf} = 0, \quad t_n < t \le t_{n+1}, \\
		& f_{pf}(\x,\bxi,t_n^+) = \beta f_u(\x,\bxi,t_n^-),
	\end{aligned}
	\right.
\end{equation}
\begin{equation}
	\label{eq:BGK_fpr}
	\left\{
	\begin{aligned}
		&\partial_t f_{pr} +\bxi\cdot \nabla f_{pr} = -\dfrac{1}{\tau}\left(f_{pr}+f_{pf}\right), \quad t_n < t \le t_{n+1}, \\
		& f_{pr}(\x,\bxi,t_n^+) = (1-\beta)f_u(\x,\bxi,t_n^-),
	\end{aligned}
	\right.
\end{equation}
\begin{equation}
	\label{eq:BGK_fpc}
	\left\{
	\begin{aligned}
		&\partial_t f_{pc} +\bxi\cdot \nabla f_{pc} = -\dfrac{1}{\tau}\left(f_{pc}-g_p\right), \quad t_n < t \le t_{n+1}, \\
		& f_{pc}(\x,\bxi,t_n^+) = 0,
	\end{aligned}
	\right.
\end{equation}
\begin{equation}
	\label{eq:BGK_fwf}
	\left\{
	\begin{aligned}
		&\partial_t f_{wf} +\bxi\cdot \nabla f_{wf} = 0, \quad t_n < t \le t_{n+1}, \\
		& f_{wf}(\x,\bxi,t_n^+) = \beta f_c(\x,\bxi,t_n^-)\approx \mathcal{E}(\beta \W_c(\x,t_n^-),\bxi),
	\end{aligned}
	\right.
\end{equation}
\begin{equation}
	\label{eq:BGK_fwr}
	\left\{
	\begin{aligned}
		&\partial_t f_{wr} +\bxi\cdot \nabla f_{wr} = -\dfrac{1}{\tau}\left(f_{wr}+f_{wf}\right), \quad t_n < t \le t_{n+1}, \\
		& f_{wr}(\x,\bxi,t_n^+) = (1-\beta)f_c(\x,\bxi,t_n^-),
	\end{aligned}
	\right.
\end{equation}
\begin{equation}
	\label{eq:BGK_fwc}
	\left\{
	\begin{aligned}
		&\partial_t f_{wc} +\bxi\cdot \nabla f_{wc} = -\dfrac{1}{\tau}\left(f_{wc}-g_w\right), \quad t_n < t \le t_{n+1}, \\
		& f_{wc}(\x,\bxi,t_n^+) = 0,
	\end{aligned}
	\right.
\end{equation}
\end{subequations}
where $g_p$ and $g_w$ are the equilibrium distribution functions of particle and wave phases,
\begin{equation}
	g_\alpha(\x,\bxi,t)=\dfrac{\rho_\alpha}{(2\pi RT)^{D/2}}\exp\left(-\dfrac{|\bxi-\u|^2}{2RT}\right)=\dfrac{\rho_\alpha}{\rho}g(\W(\x,t),\bxi),\quad \alpha=p, w
\end{equation}
with $\rho_p=\int{(f_{pf}+f_{pr}+f_{pc})\d \bxi}$ and $\rho_w=\int{(f_{wf}+f_{wr}+f_{wc})\d \bxi}$. Note that $\rho_p(t_n^\pm)=\rho_u(t_n^-)$ and $\rho_w(t_n^\pm)=\rho_c(t_n^-)$, and therefore $g_w(t_n^+)=\mathcal{E}(\W_c'(t_n^-))$ with $\W_c'=(\rho_c/\rho)\W$, as used in Eq. \eqref{eq:fwr}.

Although the kinetic equations for different types of particles/molecules are clearly presented, the UGKWP does not solve these kinetic equations separately. Instead, the time evolutions of uncollided simulation particles in the particle phase ($f_{pu}\equiv f_{pf}+f_{pr}$) and those resampled from wave phase ($f_{wf}$) are described collectively by tracking particle movement. Similarly,the time evolutions of all collided particles and molecules ($f_c=f_{pc}+f_{wc}$) are also described together deterministically. In summary, the kinetic system corresponding to the UGKWP method can be written as
\begin{subequations}
	\label{eq:4Population}
\begin{equation}
	\label{eq:WP_fpu}
	\left\{
	\begin{aligned}
		&\partial_t f_{pu} +\bxi\cdot \nabla f_{pu} = -\dfrac{1}{\tau}f_{pu}, \quad t_n < t \le t_{n+1}, \\
		& f_{pu}(\x,\bxi,t_n^+) = f_u(\x,\bxi,t_n^-).
	\end{aligned}
	\right.
\end{equation}
\begin{equation}
	\label{eq:WP_fwf}
	\left\{
	\begin{aligned}
		&\partial_t f_{wf} +\bxi\cdot \nabla f_{wf} = 0, \quad t_n < t \le t_{n+1}, \\
		& f_{wf}(\x,\bxi,t_n^+) = \beta f_c(\x,\bxi,t_n^-)\approx \mathcal{E}(\beta \W_c(\x,t_n^-),\bxi),
	\end{aligned}
	\right.
\end{equation}
\begin{equation}
	\label{eq:WP_fwr}
	\left\{
	\begin{aligned}
		&\partial_t f_{wr} +\bxi\cdot \nabla f_{wr} = -\dfrac{1}{\tau}\left(f_{wr}+f_{wf}\right), \quad t_n < t \le t_{n+1}, \\
		& f_{wr}(\x,\bxi,t_n^+) = (1-\beta)f_c(\x,\bxi,t_n^-),
	\end{aligned}
	\right.
\end{equation}
\begin{equation}
	\label{eq:WP_fc}
	\left\{
	\begin{aligned}
		&\partial_t f_{c} +\bxi\cdot \nabla f_{c} = -\dfrac{1}{\tau}\left(f_{c}-g\right), \quad t_n < t \le t_{n+1}, \\
		& f_{c}(\x,\bxi,t_n^+) = 0,
	\end{aligned}
	\right.
\end{equation}
\end{subequations}
where the kinetic equation for $f_{pu}$ is obtained by summing up Eqs. \eqref{eq:BGK_fpf}, \eqref{eq:BGK_fpr}, and \eqref{eq:BGK_fwf}, while that for $f_c$ is derived from Eqs. \eqref{eq:BGK_fpc} and \eqref{eq:BGK_fwc}.

Some comments on the above four-population kinetic system \eqref{eq:4Population} are in order.
\begin{enumerate}
\item 
The kinetic system shows clearly the underlying dynamics of four types of particles and molecules, i.e., the relaxation process of uncollided particles in particle phase (\eqref{eq:WP_fpu}), the free transport process of particles resampled from wave phase (\eqref{eq:WP_fwf}), the relaxation process of the collisional molecules  (\eqref{eq:WP_fwr}), and the collision process of all collided particles and molecules \eqref{eq:WP_fc}. The four kinetic equations are coupled through the re-initialization and relaxation terms. Specifically, discrete particles are generated from wave phase at the beginning of each time step, as described in Eq. \eqref{eq:WP_fwf}, and the collisional particles are removed from particle phase after experiencing their first collisions to become molecules in wave phase, as described by Eqs. \eqref{eq:WP_fpu} and \eqref{eq:WP_fc}. 

\item 
The system can be viewed as a collision decomposition of the original BGK equation. Specifically, the kinetic equation \eqref{eq:BGK} is decomposed into a sum of four sub-equations. This decomposition is different from the classical time-splitting method.  Actually, if $f_{pu}$, $f_{wf}$, $f_{wr}$, and $f_c$ are solutions of the corresponding sub-equations, $f=f_{pu}+f_{wf}+f_{wr}+f_c$ will satisfy the original kinetic equation \eqref{eq:BGK}. In contrast, the solutions of the sub-equations in the time-splitting system do not have this property. 

\item 
The present four-population decomposition is a generalization of the two-population one developed in Ref. \cite{ref:Coll-Hybrid-BGK2024}. In the two-population decomposition, the distribution function $f$ is separated into an uncollided part $f_u$ and a collided part $f_c$, governed by Eqs. \eqref{eq:BGK_fu} and \eqref{eq:BGK_fc}, respectively. The approach was initially developed for linear kinetic equations based on the idea of ``first-collision source" \cite{ref:Coll-Hybrid2013}, and was extended to the Boltzmann-BGK equation recently \cite{ref:Coll-Hybrid-BGK2024}. The key difference between the two-population and the present four population systems lies in the description of the uncollided particles and molecules. In the four-population system, the free-transport wave molecules ($f_{wf}$), removed collisional wave molecules ($f_{wr}$), and uncollided simulation particles ($f_{pu}$) are described separately. In contrast, all uncollided molecules and particles are indistinguishable and treated as a entity in the two-population system. In other words, the two-population system can be reduced from the four-population system by combining the three sub-equations into a single one, i.e., Eq. \eqref{eq:UGKP_fu} for $f_u=f_{wf}+f_{wr}+f_{pu}$. This means that the two-population system is just the underlying kinetic representation of the UGKP method. We highlight here that the UGKP and UGKWP methods \cite{ref:WP_LiuJCP2020} were developed independently and prior to the two-population hybrid method \cite{ref:Coll-Hybrid-BGK2024}.

\item 
Similar to the two-population system associated with the UGKP method, the four-population system can also offersome advantages for constructing numerical methods. Specifically, the linear Eqs. \eqref{eq:WP_fpu}, \eqref{eq:WP_fwf}, and \eqref{eq:WP_fwr} can be solved in relatively easy ways, while the nonlinear Eq. \eqref{eq:WP_fc} depends only on macroscopic variables and can be addressed through its moment equation. 
\end{enumerate}

Finally, we present the conservative moment equations of the kinetic system,
\begin{subequations}
\begin{equation}
\label{eq:Wpu}
    \left\{
\begin{aligned}
    &\partial_t \W_{pu} + \nabla\cdot \bm{F}_{pu} =-\dfrac{1}{\tau}\W_{pu}, \quad t_n < t \le t_{n+1},\\
    & \W_{pu}(\x,t_n^+)=\W_u(\x,t_n^-),
\end{aligned}
\right.
\end{equation}
\begin{equation}
\label{eq:Wwf}
    \left\{
\begin{aligned}
    &\partial_t \W_{wf} + \nabla\cdot \bm{F}_{wf} =0, \quad t_n < t \le t_{n+1},\\
    & \W_{wf}(\x,t_n^+)=\beta\W_c(\x,t_n^-),
\end{aligned}
\right.
\end{equation}
\begin{equation}
\label{eq:Wwr}
    \left\{
\begin{aligned}
    &\partial_t \W_{wr} + \nabla\cdot \bm{F}_{wr}  = \dfrac{1}{\tau}(\W_{wr}+\W_{wf}), \quad t_n < t \le t_{n+1},\\
    & \W_{wr}(\x,t_n^+)=(1-\beta)\W_c(\x,t_n^-),
\end{aligned}
\right.
\end{equation}
\begin{equation}
	\label{eq:Wc}
	\left\{
	\begin{aligned}
		&\partial_t \W_{c} + \nabla\cdot \bm{F}_{c}  = \dfrac{1}{\tau}(\W_{c}-\W), \quad t_n < t \le t_{n+1},\\
		& \W_{c}(\x,t_n^+)=0,
	\end{aligned}
	\right.
\end{equation}
\end{subequations}
where
\begin{equation}
\W_\alpha =\int\bpsi f_\alpha \d \bxi, \quad \bm{F}_\alpha=\int \bxi\bpsi f_\alpha \d \bxi, \qquad \alpha=pu, wf, wr, c.
\end{equation}
It is obvious that the total conservative macroscopic variables $\W=\W_{pu}+\W_{wf}+\W_{wr}+\W_h$ satisfy the conservative equation \eqref{eq:Wt}, with the flux $\bm{F}=\bm{F}_{pu}+\bm{F}_{wf}+\F_{wr}+\bm{F}_c$. The source terms and the re-initialized conditions at each time step reflect the exchange between the particle and wave phases.

\subsection{Asymptotic behaviors of the kinetic system}
\label{sec:Asymptotic}
We now analyze the asymptotic limits of the six-population kinetic system \eqref{eq:WP-6Parts}. First we give the analytical solutions of of the sub-kinetic equations,
\begin{subequations}
\label{eq:6AnaSolution}
\begin{equation}
	\label{eq:Ana_fpf}
	f_{pf}(\x,\bxi,t)=\beta f_u(\x-\bxi(t-t_n),\bxi,t_n^-), 
\end{equation}
\begin{equation}
	\label{eq:Ana_fpr}
	f_{pr}(\x,\bxi,t) =\left[e^{-(t-t_n)/\tau} - \beta \right] f_u(\x-\bxi (t-t_n),\bxi,t_n^-),
\end{equation}
\begin{equation}
	\label{eq:Ana_fpc}
	f_{pc}(\x_{ij},\bxi,t)= \dfrac{1}{\tau} \int_{t_n}^t {e^{-(t-t')/\tau} g_p(\x'_{ij},\bxi, t') \d t'}, \quad \x'_{ij}=\x_{ij}-\bxi(t-t')
\end{equation}
\begin{equation}
	\label{eq:Ana_fwf}
	f_{wf}(\x,\bxi,t)=\beta f_c(\x-\bxi(t-t_n),\bxi,t_n^-), 
\end{equation}
\begin{equation}
	\label{eq:Ana_fwr}
	f_{wr}(\x,\bxi,t) =\left[e^{-(t-t_n)/\tau} - \beta \right] f_c(\x-\bxi (t-t_n),\bxi,t_n^-),
\end{equation}
\begin{equation}
	\label{eq:Ana_fwc}
	f_{wc}(\x_{ij},\bxi,t)= \dfrac{1}{\tau} \int_{t_n}^t {e^{-(t-t')/\tau} g_w(\x'_{ij},\bxi, t') \d t'}.
\end{equation}
\end{subequations}
From the above results, we can further obtain the expressions of some reduced distribution functions,
\begin{subequations}
\begin{equation}
	\label{eq:Ana_fpu}
	f_{pu}(\x,\bxi,t) = f_{pf}+f_{pr} = e^{-(t-t_n)/\tau} f_u(\x-\bxi(t-t_n),\bxi,t_n^-),
\end{equation}
\begin{equation}
	\label{eq:Ana_fwu}
	f_{wu}(\x,\bxi,t) = f_{wf}(\x,\bxi,t) + f_{wr}(\x,\bxi,t)= e^{-(t-t_n)/\tau} f_c(\x-\bxi(t-t_n),\bxi,t_n^-),
\end{equation}
\begin{equation}
	\label{eq:Ana_fu}
	f_{u}(\x,\bxi,t) = f_{pu}(\x,\bxi,t) + f_{wu}(\x,\bxi,t) =e^{-(t-t_n)/\tau} f(\x-\bxi(t-t_n),\bxi,t_n^-),
\end{equation}
\begin{equation}
	\label{eq:Ana_fc}
	\begin{aligned}
		f_c(\x,\bxi,t) &=f_{pc}(\x,\bxi,t)+f_{wc}(\x,\bxi,t)=\dfrac{1}{\tau} \int_{t_n}^t {e^{-(t-t')/\tau} g(\x',\bxi, t') \d t'}\\
		&\approx c_1 g(\x,\bxi,t_n) +c_2 \bxi\cdot\nabla g(\x,\bxi,t_n) + c_3 \partial_t g(\x,\bxi,t_n),
	\end{aligned}
\end{equation}
\end{subequations}
where $c_1$, $c_2$, and $c_3$ are given by Eq. \eqref{eq:c1-3}. 

With the above results, we then analyze the asymptotic behaviors of the kinetic system. First, it is easy to check that
\begin{equation}
	\label{eq:deltah}
	\begin{aligned}
		\lim_{\delta_h\to 0} e^{-\delta_h} & =1, &\lim_{\delta_h \to 0} e^{-(t-t_n)/\tau} =\lim_{\delta_h \to 0} [e^{-\delta_h(t-t_{n+1})/\Delta t}] =1, \\
		\lim_{\delta_h \to \infty} e^{-\delta_h} & =0, &\lim_{\delta_h \to \infty} e^{-(t-t_n)/\tau} =\lim_{\delta_h \to \infty} [e^{-\delta_h(t-t_{n+1})/\Delta t}] =0, 
	\end{aligned}
\end{equation}
where $\delta_h=\Delta t/\tau$ is proportional to the inverse of grid Knudsen number. Therefore, in the continuum limit with $\delta_h \to \infty$, we can deduce from Eq. \eqref{eq:Ana_fu} that $f_{u}\to 0$, meaning that uncollided particles/molecules are negligible during the time interval $(t_n, t_{n+1}]$. Furthermore, it can be shown from Eq. \eqref{eq:c1-3} that $c_1\to 1$, $c_2\to -\tau$, and $c_3\to t-t_n-\tau$. Therefore, 
\begin{equation}
f_c(\x,\bxi,t)\approx g(\x,\bxi,t_n) -\tau \bxi\cdot\nabla g(\x,\bxi,t_n) + (t-t_n-\tau) \partial_t g(\x,\bxi,t_n), \end{equation}
which is exactly the Chapman-Enskog approximation of the distribution of $f(\x,\bxi,t)$ at the Navier-Stokes level. The above arguments suggest that in the continuum limit, the kinetic equations [\eqref{eq:BGK_fpc} and \eqref{eq:BGK_fwc}] for collided particles/molecules dominate the whole system, or Eq. \eqref{eq:WP_fc} dominates the four-population system \eqref{eq:4Population}. Moreover, the total distribution function $f\approx f_{c}$ recovers the Navier-Stokes equations.  

Next, we consider the free molecular limit ($\delta_h\to 0$). In this case, $\beta\to 1$ and the coefficients $c_i$'s in Eq. \eqref{eq:Ana_fc} satisfy $c_i\to 0$ ($i=1, 2, 3$). Therefore, we have  $f_c(\x,\bxi,t)\to 0$, and
\begin{equation}
\begin{aligned}
	f_{pf}(\x,\bxi,t)&\approx  f_u(\x-\bxi(t-t_n),\bxi,t_n^-),\\      
	f_{wf}(\x,\bxi,t)&\approx  f_c(\x-\bxi(t-t_n),\bxi,t_n^-).   
\end{aligned}
\end{equation}
This implies that the kinetic equations for free-transport particles and molecules, Eqs. \eqref{eq:BGK_fpf} and \eqref{eq:WP_fwf}, dominate the system, and the total distribution function $f\approx f_{pf}+f_{wf}$ recovers the collisionless BGK equation.

It is particularly interesting to examine the asymptotic behaviors of Eqs. \eqref{eq:BGK_fpr} and \eqref{eq:BGK_fwr}, which describe the removed collisional particles and wave molecules, respectively. From Eq. \eqref{eq:deltah}, we can observe that $e^{-(t-t_n)/\tau}-\beta \to 0$ in both continuum ($\delta_h\to \infty$) and collisionless ($\delta_h\to 0$) limits. Consequently, from Eqs.  \eqref{eq:Ana_fpr} and \eqref{eq:Ana_fwr} we can obtain that 
\begin{equation}
\begin{aligned}
\lim_{\delta_h\to 0}f_{pr}(\x,\bxi,t)&=\lim_{\delta_h\to \infty}f_{pr}(\x,\bxi,t) = 0, \\
\lim_{\delta_h\to 0} f_{wr}(\x,\bxi,t)&=\lim_{\delta_h\to \infty} f_{wr}(\x,\bxi,t) =0, 
\end{aligned}
\end{equation}
which suggests that the contributions from the collisional particles/molecules are negligible in both continuum and free molecular regimes. However, it should be emphasized that the contributions are significant in the transitional regime when $\delta_h$ is finite. 
This is well consistent with the definition and decomposition of the distribution functions for collisional particles and molecules. As mentioned previously, collisional particles and molecules, represented by $f_{pr}$ and $f_{wr}$, respectively, transport freely before their first collisions and are removed once collisions occur. Therefore, both $f_{pr}$ and $f_{wr}$ serve as transitional variables to describe the shift from simulation particles to wave molecules. These observations also clarify why the transitional regime is referred as ``transitional".

\subsection{Revisit UGKWP}
Now we explain the UGKWP as a hybrid numerical method based on the four-population kinetic system \eqref{eq:4Population}. The time evolution of macroscopic conservative variables in UGKWP is given by Eq. \eqref{eq:WP_tEvol}, where the cell interface flux $\F_{ij}$ is decomposed into three parts corresponding to the transports of different particles and molecules, i.e., 
\begin{equation}
	\F_{ij}=\F_{ij}^p+\F_{ij}^{wr}+\F_{ij}^c,
\end{equation}
where $\F_{ij}^p$, $\F_{ij}^{wr}$, and $\F_{ij}^c$ represent the contributions from the simulation particles, removed collisional wave molecules, and collided particles/molecules, respectively.
 $\F_{ij}^p$ is obtained by solving Eqs. \eqref{eq:WP_fpu} and \eqref{eq:WP_fwf} using the particle method, which is measured by counting the simulation particles across cell interfaces.
 $\F_{ij}^{wr}$ is calculated by solving Eq. \eqref{eq:WP_fwr} determinedly. Specifically, from the analytical solution of $f_{wr}$ given by Eq. \eqref{eq:Ana_fwr} and the approximation $f_c(t_n^-)=g_w(t_n^-)=\mathcal{E}(\W_c'(t_n^-)$, we can obtain the expression of $\F_{ij}^{wr}$ as given by Eq. \eqref{eq:fwr}. It is noted that $f_c(t_n^-)\approx g_w(t_n^-))$ is a good approximation since molecules in the wave phase experience extensive collisions and thus follow the local equilibrium state.
Finally, $\F_{ij}^c$ is calculated based on the solution of the kinetic equation \eqref{eq:WP_fc}, i.e., $f_c(\x_{ij},\bxi,t)$ given by Eq. \eqref{eq:Ana_fc}. 

The above procedure suggests that UGKWP can be viewed as a hybrid method for the kinetic system \eqref{eq:4Population}. Specifically, the kinetic equations \eqref{eq:WP_fpu} and \eqref{eq:WP_fwf} are solved using a stochastic particle method, while the kinetic equations \eqref{eq:WP_fwr} and \eqref{eq:WP_fc} are not solved directly, since only macroscopic variables are involved in the calculation. Instead, the corresponding moment equations \eqref{eq:Wwr} and \eqref{eq:Wwr} are solved deterministically. In practical implementation, UGKWP solves the overall moment equation for $\W$ rather than solving each part separately, which effectively addresses the moment equations \eqref{eq:Wwr} and \eqref{eq:Wwr}.

Furthermore, the asymptotic analysis shows that the total flux $\F_{ij}$ is consistent with the Navier-Stokes solution in continuum limit and the collisionless Boltzmann solution in free-molecular limit. Additionally, it is noted that both $\F_{ij}^{wr}$ and $\F_{ij}^c$ are evaluated from the macroscopic variables, which requires no velocity discretization. Similarly, $\F_{ij}^p$, which is evaluated from simulation particles, also avoids velocity discretization. Therefore, the UGKWP method provides an efficient multiscale solver for all Knudsen number flows.

\subsection{Two reduced three-population kinetic systems}
The time evolution of each part of the particles/molecules in UGKWP can be described by the decomposed kinetic system \eqref{eq:WP-6Parts}. Besides the four-population system given by Eq. \eqref{eq:4Population} corresponds to the implementation of UGKWP, the kinetic equations can be recombined into other forms from different viewpoints.  Here we present two reduced kinetic systems containing three populations.

First, we combine the kinetic equations for the collisonless and collisional molecules in wave phase. Specifically, we sum up Eqs. \eqref{eq:WP_fwf} and \eqref{eq:WP_fwr} to obtain the kinetic equation for the overall uncollided wave molecules. This leads to a three-population kinetic system,
\begin{subequations}
	\label{eq:3PopulationA}
	\begin{equation}
		\label{eq:WP_fpuA}
		\left\{
		\begin{aligned}
			&\partial_t f_{pu} +\bxi\cdot \nabla f_{pu} = -\dfrac{1}{\tau}f_{pu}, \quad t_n < t \le t_{n+1}, \\
			& f_{pu}(\x,\bxi,t_n^+) = f_u(\x,\bxi,t_n^-).
		\end{aligned}
		\right.
	\end{equation}
	\begin{equation}
		\label{eq:WP_fwuA}
		\left\{
		\begin{aligned}
			&\partial_t f_{wu} +\bxi\cdot \nabla f_{wu} = -\dfrac{1}{\tau}f_{wu}, \quad t_n < t \le t_{n+1}, \\
			& f_{wu}(\x,\bxi,t_n^+) =f_c(\x,\bxi,t_n^-),
		\end{aligned}
		\right.
	\end{equation}
	\begin{equation}
		\label{eq:WP_fcA}
		\left\{
		\begin{aligned}
			&\partial_t f_{c} +\bxi\cdot \nabla f_{c} = -\dfrac{1}{\tau}\left(f_{c}-g\right), \quad t_n < t \le t_{n+1}, \\
			& f_{c}(\x,\bxi,t_n^+) = 0,
		\end{aligned}
		\right.
	\end{equation}
\end{subequations}
where $f_{wu}=f_{wf}+f_{wr}$, and the total distribution is $f=f_{pu}+f_{wu}+f_c$. This system can be view as a further decomposition of the two-population system \eqref{eq:KP_2Popolation}  corresponding to the UGKP method, where $f_u$ is decomposed into $f_{pu}$ and $f_{wu}$.

Another reduced system is derived by combining the distribution functions for free transport particles and molecules ($f_{\tF}=f_{pf}+f_{wf}$), and those for the removed collisonal particles and molecules ($f_{r}=f_{pr}+f_{wr}$). The system reads 
\begin{subequations}
	\label{eq:3PopulationB}
	\begin{equation}
		\label{eq:WP_ffB}
		\left\{
		\begin{aligned}
			&\partial_t f_{\tF} +\bxi\cdot \nabla f_{\tF} = 0, \quad t_n < t \le t_{n+1}, \\
			& f_{\tF}(\x,\bxi,t_n^+) = \beta f(\x,\bxi,t_n^-).
		\end{aligned}
		\right.
	\end{equation}
	\begin{equation}
		\label{eq:WP_frB}
		\left\{
		\begin{aligned}
			&\partial_t f_{r} +\bxi\cdot \nabla f_{r} = -\dfrac{1}{\tau}(f_r+f_{\tF}), \quad t_n < t \le t_{n+1}, \\
			& f_{r}(\x,\bxi,t_n^+) =(1-\beta)f(\x,\bxi,t_n^-),
		\end{aligned}
		\right.
	\end{equation}
	\begin{equation}
		\label{eq:WP_fcB}
		\left\{
		\begin{aligned}
			&\partial_t f_{c} +\bxi\cdot \nabla f_{c} = -\dfrac{1}{\tau}\left(f_{c}-g\right), \quad t_n < t \le t_{n+1}, \\
			& f_{c}(\x,\bxi,t_n^+) = 0,
		\end{aligned}
		\right.
	\end{equation}
\end{subequations}
This system can also be viewed as a further decomposition of the two-population system \eqref{eq:KP_2Popolation}, in which the distribution function for all uncollided particles/molecules,  $f_u$, is decomposed into $f_{\tF}$ and $f_{r}$.

The asymptotic behaviors of the above two three-population kinetic systems can also be analyzed as demonstrated in Sec. \ref{sec:Asymptotic}. Moreover, the two systems can provide a basis for developing efficient kinetic schemes for flows covering a wide range of flow regimes, as mentioned in Sec. \ref{sec:BGK}.

\section{Re-interpretation of two other kinetic schemes}
With the realization that the UGKWP method can be viewed as a hybrid method based on the four-population decomposition system, we now examine two other kinetic schemes to explore thier connections with the two-population decomposition system \eqref{eq:KP_2Popolation} corresponding to the UGKP method, which can be viewed as a reduced system of the six-population one \eqref{eq:WP-6Parts}.

The first method considered is the UGKS \cite{ref:UGKS}, which solves the original kinetic equation \eqref{eq:BGK} using a discrete velocity scheme in finite-volume formulation. Specifically, the velocity space $\mathcal{R}^D$ is discretized into a discrete velocity set $\mathcal{V}_K=\{\bxi_k: \bxi_k\in \mathcal{R}^D, k=1,\cdots,K\}$, then the discrete-velocity Boltzmann-BGK equation \eqref{eq:BGK} is discretized by integrating it over a cell from $t_n$ to $t_{n+1}$,
\begin{equation}
\label{eq:UGKS-center}
    f_{k,i}^{n+1} - f_{k,i}^{n} +  \dfrac{1}{|V_i|}\sum_{j\in N(i)}\mathcal{F}_{k,ij} |S_{ij}| = \dfrac{\Delta t}{2}\left(Q_{k,i}^{n+1}+Q_{k,i}^n\right)
\end{equation}
where 
\begin{subequations}
\begin{equation}
    Q_{k,i}=-\dfrac{1}{\tau}\left[f_{k,i}-g_{k,i}\right],
\end{equation}
\begin{equation}
    f_{k,i}^n=\dfrac{1}{|V_i|}\int_{V_i}{f(\x,\bxi_k,t_n) \d\x},
\end{equation}
\begin{equation}
    \mathcal{F}_{k,ij}=\int_{t_n}^{t_{n+1}}(\bxi_k \cdot\n_{ij}){f}_k(\x_{ij},t)\d t,
\end{equation}    
\end{subequations}
with $g_{k,i}=g(\W_i^n,\bxi_k)$ and $\W_i=\sum_{k=1}^K\psi(\bxi_k) f_{k,i}$. The key point in UGKS is to determine the distribution function $f_k(\x_{ij},t)$ at cell interface, such that the numerical flux $\mathcal{F}_{ij}$ can be evaluated. This is achieved by employing the formal integral solution \eqref{eq:IntSolution} with discrete velocities $\bxi_k$, i.e, 
\begin{equation}
f_k(\x_{ij},t)=f_{u,k}(\x_{ij},t)+f_{c,k}(\x_{ij},t),    
\end{equation}
where $f_{u,k}$ and $f_{c,k}$ are the solutions of Eqs. \eqref{eq:UGKP_fu} and \eqref{eq:UGKP_fc} with the discrete velocities, respectively, namely, 
\begin{equation}
f_{u,k}(\x_{ij},t)=e^{-(t-t_n)/\tau } f(\x'_{ij},\bxi_k,t_n),
\end{equation}
\begin{equation}
f_{c,k}^h(\x_{ij},t)=c_1 g(\x'_{ij},\bxi_k,t_n) +c_2 \bxi_k\cdot\nabla g(\x'_{ij},\bxi_k,t_n) + c_3 \partial_t g(\x'_{ij},\bxi_k,t_n),
\end{equation}
with $\x'_{ij}=\x_{ij}-\bxi_k(t-t_n)$.  Note that no resampling is required in the UGKS since the discrete distribution functions are tracked in the evolution. From this point of view, the UGKS actually solves the two-population kinetic system \eqref{eq:KP_2Popolation}.

The second kinetic method considered is the DUGKS \cite{ref:DUGKS13}, which is also a finite-volume discrete-velocity method similar to UGKS. The evolution of cell-averaged distribution function in DUGKS is the same as Eq. \eqref{eq:UGKS-center}, but the distribution function at cell interface is evaluated in a simpler way. Specifically, the mid-point rule is used to approximate the time integration in the flux,
\begin{equation}
    \mathcal{F}_{k,ij}=\int_{t_n}^{t_{n+1}}(\bxi_k \cdot\n_{ij}){f}_k(\x_{ij},t)\d t\approx \Delta t(\bxi_k \cdot\n_{ij}){f}_k(\x_{ij},t_n+\Delta t/2).
\end{equation}    
Then the distribution function at half time step is obtained by integrating Eq. \eqref{eq:BGK} along the characteristic long with a half time step,
\begin{equation}
\label{eq:DUGKS-f}
f_k(\x_{ij},t_n+\Delta t/2)-f_k(\x'_{ij}, t_n)=\dfrac{\Delta t}{4}\left[Q_k(\x_{ij},t_n+\Delta t/2)+Q_k(\x'_{ij}, t_n)\right],
\end{equation}
where $\x'_{ij}=\x_{ij}-\bxi_k \Delta t/2$ and the trapezoidal rule is applied to the time integration of the collision term. In order to reveal the connection with the two-population decomposition system, we apply the same integration to Eqs. \eqref{eq:UGKP_fu} and \eqref{eq:UGKP_fc} to obtain
\begin{equation}
\label{eq:DUGKS-fu}
f_{u,k}(\x_{ij},t_n+\Delta t/2)-f_{p,k}(\x'_{ij}, t_n)=-\dfrac{\Delta t}{4\tau}\left[f_{p,k}(\x_{ij},t_n+\Delta t/2)+f_{p,k}(\x'_{ij}, t_n)\right],
\end{equation}
and
\begin{equation}
\label{eq:DUGKS-fc}
\begin{aligned}
f_{c,k}(\x_{ij},t_n+\Delta t/2)-f_{h,k}(\x'_{ij}, t_n) & = -\dfrac{\Delta t}{4\tau}\left[f_{h,k}(\x_{ij},t_n+\Delta t/2)-g_k(\x_{ij},t_n+\Delta t/2)\right]\\
& \quad -\dfrac{\Delta t}{4\tau}\left[f_{h,k}(\x'_{ij}, t_n)-g_k(\x'_{ij}, t_n)\right].   
\end{aligned}
\end{equation}
It can be seen that Eq. \eqref{eq:DUGKS-f} is exactly the summution of Eqs. \eqref{eq:DUGKS-fu} and \eqref{eq:DUGKS-fc}, suggesting that DUGKS can also be viewed as a numerical method based on the two-population decomposition system.

\section{Summary}
The UGKWP is a recently developed numerical method for multiscale flows within the framework of direct modeling at discrete level, which separates the gas system into particle phase and hydrodynamic wave phase. The particle phase consists of the original simulation particles and those resampled from a portion of the wave, while the wave phase is composed of real gas molecules. Based on the collision process and phase state, particles and wave molecules are further classified into six types: collisionless particles/molecules transporting freely, collisional particles/molecules experiencing free flight and collisions that will be removed once a collision occurs, and collided particles/molecules experiencing extensive collisions. To understand the underlying kinetics of UGKP, kinetic equations are presented for the different types of particles/molecules. 

The resulting kinetic system can be viewed as a six-population decomposition of the original kinetic equation with re-initialization, which is then recombined into a four-population system corresponding to the practical implementation of UGKWP. Specifically, the four-population system is composed of kinetic equations for all the original uncollided particles ($f_{pu}=f_{pf}+f_{pr}$), collisionless wave molecules ($f_{wf}$), collisional wave molecules ($f_{wr}$), and collided particles/molecules ($f_{c}=f_{pc}+f_{wc}$). Additionally, two reduced three-polulation kinetic systems are provided by different recombinations.  

The asymptotic behaviors of the six-population kinetic system are also analyzed. It is shown that the kinetic equations for uncollided particles/molecules dominate the system in free-molecular regime, while the kinetic equations for the collided particles/molecules dominate the system in the continuum limit. Interestingly, the kinetic equations for the collisional particles/molecules play negligible roles in both limits.

Based on the four-population kinetic system, the UGKWP is revisited as a collision hybrid method. Additionally, two other kinetic schemes (UGKS and DUGKS) are analyzed to demonstrate their connections with the reduced two-population kinetic system. The results suggest that the decomposition system and reduced systems can also serve as a basis for developing new kinetic methods for multiscale flows.

\acknowledgments{This article is dedicated to the memory of Prof. Jiequan Li, a dear friend and
collaborator of the authors. Z.L.G. is supported by the Interdiciplinary Research Program of HUST (2023JCYJ002), and part of this work was carried out during his visit to the Department of Mathematics at the Hong Kong University of Science and Technology. K.X. is supported by Hong Kong research grant council (16208021, 16301222).}
\bibliography{ReWP}

\end{document}